\documentclass[aps,prl,superscriptaddress,twocolumn,floatfix,showpacs,amsmath]{revtex4}
\usepackage{graphicx}
\usepackage{latexsym}
\usepackage{psfrag}
\usepackage{dcolumn}

\begin{document}

\title{Ferromagnetic-like closure domains in ferroelectric ultrathin films.}

\author{ Pablo Aguado-Puente }
\affiliation{ CITIMAC, Universidad de Cantabria, 
              Avda. de los Castros s/n, E-39005 Santander, Spain}
\author{ Javier Junquera }
\affiliation{ CITIMAC, Universidad de Cantabria, 
              Avda. de los Castros s/n, E-39005 Santander, Spain}
\date{\today}

\begin{abstract}
 We simulate from first-principles the energetic, structural, and electronic
 properties of 
 ferroelectric domains in ultrathin capacitors made of 
 a few unit cells of BaTiO$_3$ between two 
 metallic SrRuO$_3$ electrodes in short circuit.
 The domains are stabilized down 
 to two unit cells, adopting the form of a domain of closure, common 
 in ferromagnets but only recently detected experimentally in 
 ferroelectric thin films.
 The domains are closed by the in-plane relaxation of the atoms in the first
 SrO layer of the electrode, that behaves more like SrO in highly polarizable
 SrTiO$_3$ than in metallic SrRuO$_3$.
 Even if small, these lateral displacements are essential to stabilize the
 domains, and might provide some hints to explain why some systems 
 break into domains while others remain in a monodomain configuration. 
 An analysis of the electrostatic potential 
 reveals preferential points of pinning for charged defects
 at the ferroelectric-electrode interface,
 possibly playing a major role in films fatigue. 
\end{abstract}

\pacs{77.80.Dj,77.22.Ej,77.84.Dy,68.55.-a}
 
\maketitle

 Ultrathin film ferroelectric capacitors are under active
 investigation \cite{Dawber-05, Ghosez-06}. 
 Of considerable technological interest as
 memories, transducers, and electromechanical devices, 
 they present problems of considerable scientific interest.
 Although technologically relevant films are thicker than 100 nm, 
 deeper understanding of the origin of these problems 
 requires combined experimental and theoretical studies of thinner regimes.
 On the one hand, recent breakthroughs on materials synthesis 
 and characterization 
 techniques have allowed the growth of ferroelectric thin films
 with a control at the atomic scale and the local measurement of the 
 ferroelectric properties \cite{Ahn-04}.
 On the other, 
 the steady increase in computational power and improvements in the 
 efficiency of the algorithms permit accurate first-principles 
 study of larger and
 more complex systems, overlapping in size with
 those grown epitaxially.
 
 Prominent among problems of interest 
 is understanding the mechanisms screening charge
 densities at the interfaces.
 The termination of the ferroelectric polarization at the 
 surface or the electrode interface 
 generates a polarization charge which
 gives rise to a depolarizing field tending to suppress the polarization.
 Two mechanisms are
 traditionally invoked for the compensation of the polarization charges:
 the first, 
 screening by charge accumulation at the electrode
 (or even by ionic adsorbates \cite{Fong-06,Spanier-06});
 the second, 
 the breaking up of the system into 
 domains \cite{Streiffer-02,Fong-04}.

 Previous first principles local density calculations on 
 realistic short-circuited ferroelectric capacitors
 suggested that a 
 monodomain configuration for the polarization was unstable below a 
 critical thickness that ranged between $m=2$ 
 and $m=6$ layers \cite{Junquera-03.1,Umeno-06,Duan-06} 
 of ferroelectric, depending on the perovskite,
 the electrode,
 and the termination at the interface.
 In all these approaches the electrode was the only source of screening,
 providing free charges that accumulate
 at the interface on the metallic side and even decay exponentially
 into the first few layers of the ferroelectric, or sharing the 
 ionic displacements responsible for the polarization 
 in the ferroelectric \cite{Gerra-06}.
 In any case, the mechanism is ineffective 
 below this critical thickness where the paraelectric phase was stabilized.

\begin{figure}[htbp]
    \begin{center}
       \includegraphics[width=\columnwidth] {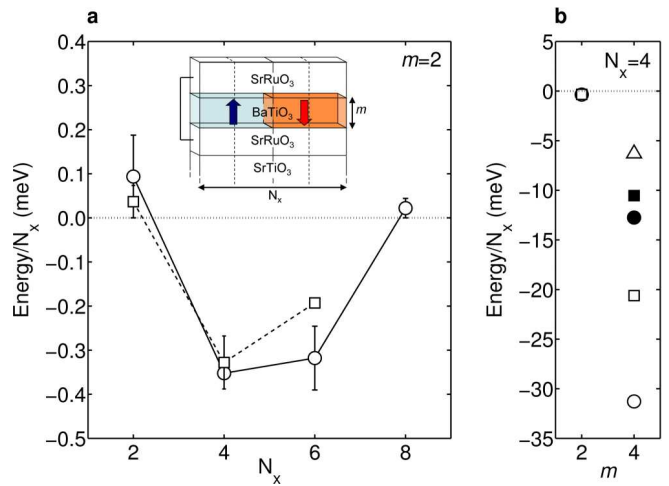}       
       \caption{ (color online).
                 Difference in energy between polydomain and 
                 paraelectric phases as a function of 
                 (a) the domain period $N_{x}$
                 for a ferroelectric thin film two unit cells thick ($m=2$),
                 and (b)
                 the thickness of the ferroelectric film
                 for a capacitor with $N_{x} = 4$.
                 The energy of the paraelectric phase 
                 (dotted line) is taken as reference.
                 First-principles results for both 
                 Ba-centred (circles, solid line)
                 and Ti-centred (squares, dashed) domain walls are shown.
                 In (a) differences in energies between local minima 
                 of the polydomain
                 phase are represented by error bars.
                 Inset, structure of the ferroelectric capacitor considered. 
                 $N_{x}$ is the stripe period and $m$ is the 
                 thickness of the ferroelectric thin film, in  
                 number of unit cells of the ferroelectric perovskite oxide.
                 In (b) the result for the most stable monodomain configuration
                 is also shown (triangle). 
                 Full symbols correspond to constrained relaxations 
                 where no in-plane displacements are allowed.
               }
       \label{fig:energy}
    \end{center}
 \end{figure}

 In this letter we simulate from first principles, 
 within the local density approximation to the density functional theory
 and the numerical atomic orbital 
 method as implemented in the {\sc Siesta} code \cite{Soler-02}, 
 typical SrRuO$_3$/BaTiO$_3$/SrRuO$_3$ ferroelectric capacitors, in which
 we allow the system to form
 domains. 
 Our starting point is the reference paraelectric heterostructure
 described in Ref. \cite{Junquera-03.1}, that is now replicated
 $N_{x}$ times along the [100] direction,
 where $N_{x}$ ranges from 2 to 8.
 A soft mode distortion of the bulk tetragonal phase is superimpossed
 to the BaTiO$_3$ layers of the previous paraelectric configuration, so
 the polarization points upwards in half of the superlattice and downwards 
 in the other half [see inset of Fig. \ref{fig:energy}(a)].
 The twinning on both the BaO (Ba-centred), and TiO$_2$ (Ti-centred)
 planes is considered.
 Then, the atomic positions of all the ions, both in the electrode and in the
 ferroelectric thin film, are relaxed till the maximum component
 of the force on any atom is smaller than 0.01 eV/\AA\ for $m=2$,
 and 0.04 eV/\AA\ for $m=4$.
 Very accurate computations are required since the differences in energy
 between relevant phases are eight orders of magnitude smaller than
 the absolute value of the energy.
 The electronic density, Hartree, and exchange-correlation potentials are
 computed in a uniform real space grid, with an 
 equivalent plane-wave cutoff of 400 Ry.
 Once self-consistency is achieved,
 the grid is refined (reducing the distance between points by half) to
 compute the total energy and atomic forces.
 We used a $N_{k_{x}} \times 12 \times 1$ Monkhorst-Pack mesh for all
 the Brillouin zone integrations, where $N_{k_{x}} = \frac{12}{N_{x}}$
 except for the interface with $N_{x}$ =8, where $N_{k_{x}}$ = 2.
 All the calculations are performed at T = 0.
 Details on pseudopotentials and basis set used can be found 
 in Ref. \cite{Junquera-03.2}.

 Our calculations support stabilization of a polydomain phase
 with an exceptionally small periodicity
 below the previous critical thickness [see Fig. \ref{fig:energy}(a)],
 in good agreement with the results obtained with Landau 
 theory \cite{Bratkovsky-06.1}.
 For a two unit cell thick film, $m=2$, the extra source of screening 
 is efficient provided that the
 domain period is between two and 
 four times the thickness of the film. 
 Within this region, the energy cost of forming the domain wall is 
 compensated by reduction
 of the net polarization charge at the interfaces. 
 As in 180$^\circ$ stripe-domains 
 in bulk \cite{Meyer-02}, 
 the Ba-centred wall configuration is preferred. 
 The energy difference between the most stable polydomain and
 the paraelectric phase for a capacitor with $m=2$
 is very small, of the order of
 1.5 meV ($\simeq$ 16 K) for the whole supercell.
 For this thickness there is essentially no 
 energy difference between domains of lateral 
 periods $N_{x}$ = 4 and 6,
 suggesting that both might be equally present in a sample.
 Heating or cooling processes might help the system to overcome
 potential energy barriers and activate the transition between them.
 Although the conductive nature of the substrate is different, this fact might
 provide an extra source of explanation \cite{Prosandeev-07} for 
 the intriguing richness in behaviour of the stripe domain patterns
 observed experimentally in PbTiO$_3$ thin films grown on SrTiO$_3$, 
 where two different periods coexisted
 \cite{Fong-04}.  
 (Note that our ratio between domain periods,
 1.5, is close to the experimental factor 1.4 for the so-called
 $\alpha$ and $\beta$ phases in Ref. \onlinecite{Fong-04}.)

 The energy differences between polydomain and paraelectric phases increase
 very quickly with thickness [Fig.  \ref{fig:energy}(b)]
 and amounts to 120 (80) meV for a Ba-centred
 (Ti-centred) domain wall capacitor with $N_{x} = 4$ and $m = 4$.
 For this size, the polydomain phases are more stable
 than the monodomain configuration, itself more stable than the paraelectric
 phase by 20 meV.

\begin{figure}[htbp]
    \begin{center}
       \includegraphics[width=\columnwidth] {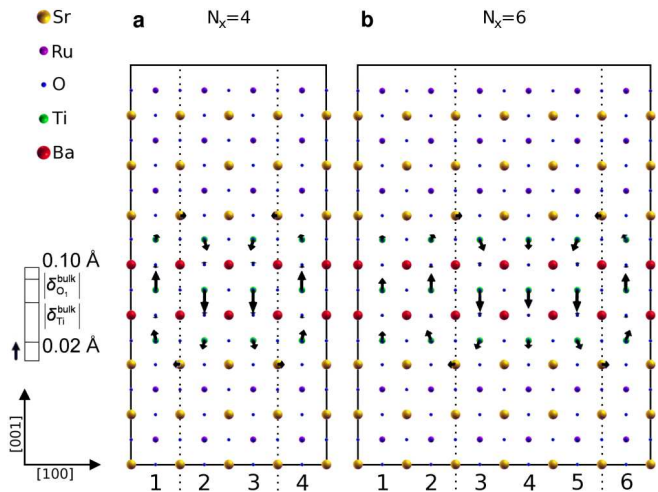}       
       \caption{ (color online). 
                 Schematic representation of the atomic relaxations 
                 in patterns of
                 domains of closure with domain period of $N_{x} = 4$ (a),
                 and $N_{x} = 6$ (b). 
                 Balls, representing atoms, are located at the positions of 
                 the reference paraelectric phase.
                 Atomic displacements for the polydomain
                 configuration after relaxation are represented by arrows,
                 whose magnitude can be gauged with respect to the 
                 displacements in the bulk tetragonal phase of BaTiO$_3$
                 at the scale on the left.
                 Dotted lines indicate the position of the domain wall. 
                 Only Ba-centred domains are shown. Similar results are
                 obtained for Ti-centred domains.
               }
       \label{fig:structure}
    \end{center}
 \end{figure}

 The minimum energy structures
 of these ferroelectric capacitors,
 shown in Fig. \ref{fig:structure}, display the
 closure domain configuration proposed
 by Landau and Lifshitz \cite{Landau-35} and Kittel \cite{Kittel-46} 
 for magnetic systems.
 At the centre of the BaTiO$_3$ layer, the displacement of the atoms
 and therefore the corresponding local dipoles, point 
 normal to the interface (coordinate $z$), 
 as expected for 180$^\circ$ stripe domains.
 However, approaching the ferroelectric/electrode interface
 a small tilt towards [100] is observed. 
 Remarkably, the domains are not closed by the surface layer of the 
 ferroelectric \cite{Lai-06} but by the 
 in-plane displacements of the Sr and O atoms at the first layer 
 of the electrode, which yield
 a closure domain pattern, with 90$^\circ$ 
 domain walls with the $z$ oriented domains inside the film.
 In contrast to the metallic relaxations in monodomain configurations,
 where ionic displacements penetrate into the metal over a distance
 of two or three unit cells \cite{Gerra-06,Stengel-06},
 the displacements beyond the 
 second RuO$_2$ layer are negligible, an indication of 
 more effective screening produced by the domains of closure.
 The in-plane displacements of the atoms at the
 interfacial SrO layer, although small in magnitude,
 stabilize the domain structure.
 If a constrained relaxation is performed in which the in-plane forces 
 on all the atoms
 are artificially eliminated, the atoms move back to the 
 paraelectric positions for $m=2$, or to a structure
 comparable in energy to the most stable monodomain configuration for
 $m=4$ [Fig. \ref{fig:energy}(b)]. 
 Whether the in-plane displacement is allowed or not might partially
 explain for the very different configurations found 
 experimentally in related heterostructures:
 Lichtensteiger {\it et al.}, using the same experimental setup,
 have observed how high-quality ultrathin films of PbTiO$_3$ 
 grown on Nb-SrTiO$_3$ electrodes remain in a 
 monodomain configuration \cite{Lichtensteiger-05}
 (although with reduced polarization and tetragonality)
 whereas they form domains when the electrode
 is replaced by 
 La$_{0.67}$Sr$_{0.33}$MnO$_3$ \cite{Lichtensteiger-07}. 
 The same domain formation is suggested for 
 Pb(Zr$_{0.2}$Ti$_{0.8}$)O$_3$ on SrRuO$_3$ \cite{Nagarajan-06}.

 Regarding the origin of this polarization induced relaxation,
 the analysis of the projected density of states 
 (not shown here) shows that
 the SrO layer closest to the interface behaves more like
 SrO in SrTiO$_3$ than SrO
 in metallic SrRuO$_3$. Similar behaviour was found in AO/ATiO$_3$ 
 heterostructures \cite{Junquera-03.2}, where A = Ba or Ti.
 Both first-principles computations \cite{Neaton-03} and 
 experimental measurements \cite{Tian-06} have shown that
 SrTiO$_3$ is highly polarizable when combined with BaTiO$_3$ in 
 heterostructures.

 Similar domain patterns have been found using a first-principles 
 effective hamiltonian for Pb(Zr$_{0.4}$Ti$_{0.6}$)O$_3$ 
 \cite{Prosandeev-07} asymmetrically screened
 (grown on a nonconducting substrate and with a metal with a dead layer 
 as top electrode), 
 and using a Landau-Ginzburg phenomenological approach for a PbTiO$_3$ thin
 film \cite{Stephenson-06}, both asymmetrically and symmetrically coated
 with insulating SrTiO$_3$.
 Here, the domains of closure are obtained 
 even for a symmetrical metal/ferroelectric/metal capacitor,
 where the metallic plates should provide significant screening.
 
\begin{figure}[htbp]
    \begin{center}
       \includegraphics[width=\columnwidth] {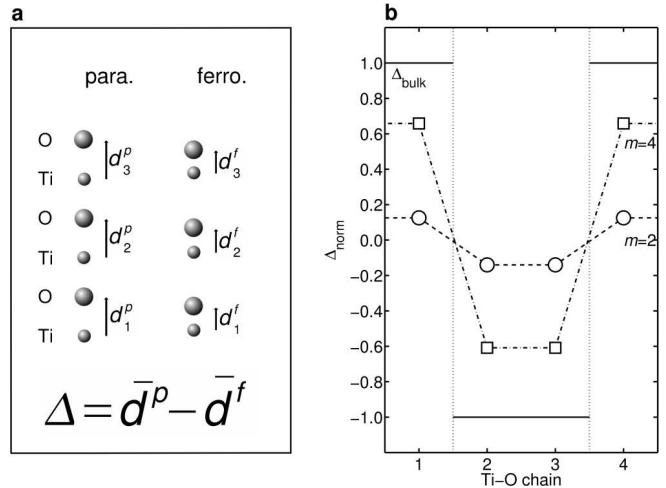}       
       \caption{ Measurement of the polarization in the ferroelectric layer
                 as a function of position along the [100] direction of
                 the capacitor. 
                 (a) Definition of the average change in distance $\Delta$
                 between Ti and apical O in a chain along [001] 
                 for an interface with $m=2$.
                 In every case, the atomic positions correspond to the
                 lowest energy structure.
                 A positive value of $\Delta$ means a polarization pointing
                 upward.
                 (b) Profile of the normalized averaged change 
                 in distance along
                 $z$ as a function of the position of the chain 
                 for a Ba-centred interface of domain period $N_{x} = 4$. 
                 The chains are numbered as indicated in 
                 Fig. \ref{fig:structure}.
                 Results are shown for $m=2$ (dashed line) 
                 and $m=4$ (dot-dashed). 
                 Dotted lines represent the position of the domain walls.
                 }
       \label{fig:polarization}
    \end{center}
 \end{figure}

 The polarization can be estimated from the structural calculations. 
 Figure \ref{fig:polarization}  
 displays how much the polar distortion along $z$ is changed
 by the presence of a domain pattern. 
 We define as $\Delta$ the average of the change of distance,
 with respect the most stable paraelectric configuration,
 between a Ti atom and the nearest O atom lying on top along the 
 $z$ direction (cf. Ref. \onlinecite{Meyer-02}),
 normalized with respect to the short Ti-O distance
 in the tetragonal bulk phase.
 $\Delta_{norm}$ is a very sensitive indicator of the 
 polar order:
 it is zero as long as the atoms lie in the paraelectric position and 
 tends to unity as the full bulk polar distortion is attained.
 Figure \ref{fig:polarization} shows a very narrow 180$^\circ$ domain wall,
 about a lattice constant wide, across which the polar
 distortion symmetrically reverses its sign. 
 In contrast to 180$^\circ$ domains in bulk \cite{Meyer-02}, 
 where the ferroelectric distortion
 fully recovers its bulk value by the second atomic plane far away from the
 domain wall, here it only amounts to 13\% of the bulk value at the 
 centre of each domain for a $m=2$ structure, suggesting that the polarization
 for the thin film is one order of magnitude smaller than in bulk.
 This mean polarization increases with thickness, and
 already amounts to 60\% for a thin film  four unit cells thick ($m=4$). 

\begin{figure}[htbp]
    \begin{center}
       \includegraphics[width=\columnwidth] {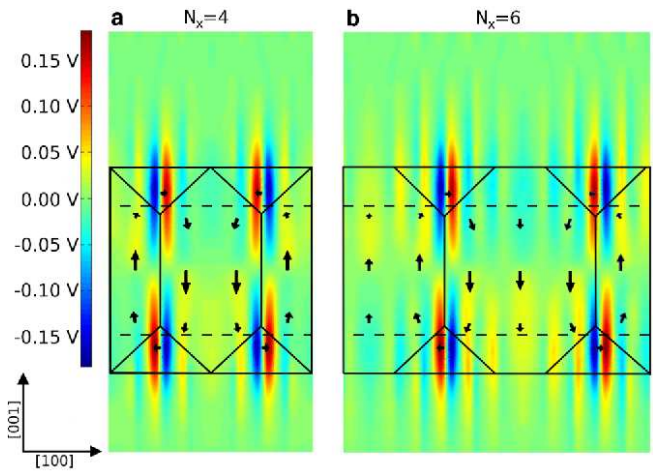}       
       \caption{ (color online).
                 Map of the nanosmoothed electrostatic potential 
                 in a two unit cells thick
                 ferroelectric capacitor with a stripe period 
                 of $N_{x} = 4$ (a),
                 and $N_{x} = 6$ (b).
                 The arrows represent the atomic displacements with respect
                 the paraelectric phase as in Fig. \ref{fig:structure}.
                 Only the displacements of the cations are shown for simplicity.
                 Full lines are a schematic representation of the 
                 domains of closure, while 
                 dashed lines mark the position of the BaTiO$_3$/SrRuO$_3$
                 interface.
               }
       \label{fig:potential}
    \end{center}
 \end{figure}

 Ideally, closure domains do not produce any polarization charge since the
 normal component of the polarization is preserved across any domain wall.
 Therefore the depolarizing field should vanish everywhere \cite{Kittel-46},
 and a constant electrostatic potential is expected.
 To further check this point we plot in Fig. \ref{fig:potential} 
 the nanosmoothed \cite{Baldereschi-88,Junquera-07} 
 electrostatic potential along $z$ as a function of the
 position along the [100] direction of the capacitor.
 No nanosmoothing is performed along $x$.
 For a stripe of thickness $m=2$ and period $N_{x} = 4$,
 the potential is esentially flat at the centre of the domain, in contrast
 to the depolarizing field reported for monodomain configurations
 \cite{Junquera-03.1}. 
 A large microscopic field along [100] 
 appears inside the domains of closure at the metal-ferroelectric interface.
 The origin of this field is due to the difference in polarization in 
 the domain of closure (the polarization along $x$ equals 0.9
 and 3.3 $\mu$C/cm$^2$ for $m$ = 2 and $m$ = 4, respectively)
 and inside the thin film in our realistic capacitor.
 Besides, after nanosmoothing in $z$
 a residual depolarizing field along [001] is identified in 
 the neighborhood of the domain wall, decaying rapidly away
 from it. This last field might be responsible for the lowering of the 
 polarization with respect to bulk shown in Fig. \ref{fig:polarization}.
 Both fields might play an important role in the fatigue of ferroelectric 
 capacitors, the most serious device problem in ferroelectric
 thin films \cite{Dawber-05}. In particular we identify at the 
 ferroelectric/electrode interface the preferred points of migration
 of charged defects, which pin the domain walls and inhibit 
 their motion \cite{He-03}.
 The depolarizing field at the centre of the domain increases with the
 domain period; it starts to be appreciable for $N_{x} = 6$ [Fig.
 \ref{fig:potential}(b)] and finally desestabilizes the 
 ferroelectric distortions
 for $N_{x} = 8$, as shown in Fig. \ref{fig:energy}(a).  

 Although we have demonstrated that the domains are stable,
 it is not clear whether
 the capacitor as a whole can be called ferroelectric since,
 for this, the polarization has to be 
 switchable under external electric fields \cite{Nagarajan-06,Lai-06}.

 Our calculations provide insightful results on the
 energetic, structural and electronic properties of 
 ferromagnetic-like closure domains in ultrathin capacitors,
 only recently observed experimentally by J. F. Scott's group 
 \cite{Scott-porto}.
 We provide some hints to explain why some systems break into domains
 while others remain in a monodomain configuration.
 We also predict the preferential sites for pinning charged defects,
 important for understanding the fatigue of thin films.

 We thank M. H. Cohen for the critical reading of the manuscript. 
 Calculations were performed on the computers
 of the ATC group and on the Altamira Supercomputer 
 at the Universidad de Cantabria.
 This work was supported by the Spanish MEC under Projects FIS2006-02261 and
 FPU AP2006-02958, 
 and the Australian Research Council ARC Discovery Grant DP 0666231.


\begin{thebibliography}{31}
\expandafter\ifx\csname natexlab\endcsname\relax\def\natexlab#1{#1}\fi
\expandafter\ifx\csname bibnamefont\endcsname\relax
  \def\bibnamefont#1{#1}\fi
\expandafter\ifx\csname bibfnamefont\endcsname\relax
  \def\bibfnamefont#1{#1}\fi
\expandafter\ifx\csname citenamefont\endcsname\relax
  \def\citenamefont#1{#1}\fi
\expandafter\ifx\csname url\endcsname\relax
  \def\url#1{\texttt{#1}}\fi
\expandafter\ifx\csname urlprefix\endcsname\relax\def\urlprefix{URL }\fi
\providecommand{\bibinfo}[2]{#2}
\providecommand{\eprint}[2][]{\url{#2}}

\bibitem[{\citenamefont{Dawber et~al.}(2005)\citenamefont{Dawber, Rabe, and
  Scott}}]{Dawber-05}
\bibinfo{author}{\bibfnamefont{M.}~\bibnamefont{Dawber}},
  \bibinfo{author}{\bibfnamefont{K.~M.} \bibnamefont{Rabe}}, \bibnamefont{and}
  \bibinfo{author}{\bibfnamefont{J.~F.} \bibnamefont{Scott}},
  \bibinfo{journal}{Rev. Mod. Phys.} \textbf{\bibinfo{volume}{77}},
  \bibinfo{pages}{1083} (\bibinfo{year}{2005}).

\bibitem[{\citenamefont{Ghosez and Junquera}(2006)}]{Ghosez-06}
\bibinfo{author}{\bibfnamefont{Ph.}~\bibnamefont{Ghosez}} \bibnamefont{and}
  \bibinfo{author}{\bibfnamefont{J.}~\bibnamefont{Junquera}},
  \emph{\bibinfo{title}{Handbook of theoretical and computational
  nanotechnology}} (\bibinfo{publisher}{American Scientific Publishers},
  \bibinfo{address}{Stevenson Ranch, CA}, \bibinfo{year}{2006}),
  vol.~\bibinfo{volume}{9}, pp. \bibinfo{pages}{623--728}.

\bibitem[{\citenamefont{Ahn et~al.}(2004)\citenamefont{Ahn, Rabe, and
  Triscone}}]{Ahn-04}
\bibinfo{author}{\bibfnamefont{C.~H.} \bibnamefont{Ahn}},
  \bibinfo{author}{\bibfnamefont{K.~M.} \bibnamefont{Rabe}}, \bibnamefont{and}
  \bibinfo{author}{\bibfnamefont{J.-M.} \bibnamefont{Triscone}},
  \bibinfo{journal}{Science} \textbf{\bibinfo{volume}{303}},
  \bibinfo{pages}{488} (\bibinfo{year}{2004}).

\bibitem[{\citenamefont{Fong et~al.}(2006)\citenamefont{Fong, Kolpak, Eastman,
  Streiffer, Fuoss, Stephenson, Thompson, Kim, Choi, Eom et~al.}}]{Fong-06}
\bibinfo{author}{\bibfnamefont{D.~D.} \bibnamefont{Fong}}
  \bibnamefont{et~al.}, 
  \bibinfo{journal}{Phys. Rev. Lett.}
  \textbf{\bibinfo{volume}{96}}, \bibinfo{pages}{127601}
  (\bibinfo{year}{2006}).

\bibitem[{\citenamefont{Spanier et~al.}(2006)\citenamefont{Spanier, Kolpak,
  Urban, Grinberg, Ouyang, Yun, Rappe, and Park}}]{Spanier-06}
\bibinfo{author}{\bibfnamefont{J.~E.} \bibnamefont{Spanier}}
  \bibnamefont{et~al.},
  \bibinfo{journal}{Nano Lett.} \textbf{\bibinfo{volume}{6}},
  \bibinfo{pages}{735} (\bibinfo{year}{2006}).

\bibitem[{\citenamefont{Streiffer et~al.}(2002)\citenamefont{Streiffer,
  Eastman, Fong, Thompson, Munkholm, Murty, Auciello, Bai, and
  Stephenson}}]{Streiffer-02}
\bibinfo{author}{\bibfnamefont{S.~K.} \bibnamefont{Streiffer}}
  \bibnamefont{et~al.},
  \bibinfo{journal}{Phys. Rev. Lett.} \textbf{\bibinfo{volume}{89}},
  \bibinfo{pages}{067601} (\bibinfo{year}{2002}).

\bibitem[{\citenamefont{Fong et~al.}(2004)\citenamefont{Fong, Stephenson,
  Streiffer, Eastman, Auciello, Fuoss, and Thompson}}]{Fong-04}
\bibinfo{author}{\bibfnamefont{D.~D.} \bibnamefont{Fong}}
  \bibnamefont{et~al.},
  \bibinfo{journal}{Science} \textbf{\bibinfo{volume}{304}},
  \bibinfo{pages}{1650} (\bibinfo{year}{2004}).

\bibitem[{\citenamefont{Junquera and Ghosez}(2003)}]{Junquera-03.1}
\bibinfo{author}{\bibfnamefont{J.}~\bibnamefont{Junquera}} \bibnamefont{and}
  \bibinfo{author}{\bibfnamefont{Ph.}~\bibnamefont{Ghosez}},
  \bibinfo{journal}{Nature (London)} \textbf{\bibinfo{volume}{422}},
  \bibinfo{pages}{506} (\bibinfo{year}{2003}).

\bibitem[{\citenamefont{Umeno et~al.}(2006)\citenamefont{Umeno, Meyer,
  Els{\"a}sser, and Gumbsch}}]{Umeno-06}
\bibinfo{author}{\bibfnamefont{Y.}~\bibnamefont{Umeno}},
  \bibinfo{author}{\bibfnamefont{B.}~\bibnamefont{Meyer}},
  \bibinfo{author}{\bibfnamefont{C.}~\bibnamefont{Els{\"a}sser}},
  \bibnamefont{and} \bibinfo{author}{\bibfnamefont{P.}~\bibnamefont{Gumbsch}},
  \bibinfo{journal}{Phys. Rev. B} \textbf{\bibinfo{volume}{74}},
  \bibinfo{pages}{060101(R)} (\bibinfo{year}{2006}).

\bibitem[{\citenamefont{Duan et~al.}(2006)\citenamefont{Duan, Sabirianov, Mei,
  Jaswal, and Tsymbal}}]{Duan-06}
\bibinfo{author}{\bibfnamefont{C.-G.} \bibnamefont{Duan}},
  \bibinfo{author}{\bibfnamefont{R.~F.} \bibnamefont{Sabirianov}},
  \bibinfo{author}{\bibfnamefont{W.-N.} \bibnamefont{Mei}},
  \bibinfo{author}{\bibfnamefont{S.~S.} \bibnamefont{Jaswal}},
  \bibnamefont{and} \bibinfo{author}{\bibfnamefont{E.~Y.}
  \bibnamefont{Tsymbal}}, \bibinfo{journal}{Nano Lett.}
  \textbf{\bibinfo{volume}{6}}, \bibinfo{pages}{483} (\bibinfo{year}{2006}).

\bibitem[{\citenamefont{Gerra et~al.}(2006)\citenamefont{Gerra, Tagantsev,
  Setter, and Parlinski}}]{Gerra-06}
\bibinfo{author}{\bibfnamefont{G.}~\bibnamefont{Gerra}},
  \bibinfo{author}{\bibfnamefont{A.~K.} \bibnamefont{Tagantsev}},
  \bibinfo{author}{\bibfnamefont{N.}~\bibnamefont{Setter}}, \bibnamefont{and}
  \bibinfo{author}{\bibfnamefont{K.}~\bibnamefont{Parlinski}},
  \bibinfo{journal}{Phys. Rev. Lett.} \textbf{\bibinfo{volume}{96}},
  \bibinfo{pages}{107603} (\bibinfo{year}{2006}).

\bibitem[{\citenamefont{Soler et~al.}(2002)\citenamefont{Soler, Artacho, Gale,
  Garc\'{\i}a, Junquera, Ordej\'on, and S\'anchez-Portal}}]{Soler-02}
\bibinfo{author}{\bibfnamefont{J.~M.} \bibnamefont{Soler}}
  \bibnamefont{et~al.},
  \bibinfo{journal}{J. Phys.: Condens. Matter} \textbf{\bibinfo{volume}{14}},
  \bibinfo{pages}{2745} (\bibinfo{year}{2002}).

\bibitem[{\citenamefont{Junquera et~al.}(2003)\citenamefont{Junquera, Zimmer,
  Ordej\'on, and Ghosez}}]{Junquera-03.2}
\bibinfo{author}{\bibfnamefont{J.}~\bibnamefont{Junquera}},
  \bibinfo{author}{\bibfnamefont{M.}~\bibnamefont{Zimmer}},
  \bibinfo{author}{\bibfnamefont{P.}~\bibnamefont{Ordej\'on}},
  \bibnamefont{and} \bibinfo{author}{\bibfnamefont{Ph.}~\bibnamefont{Ghosez}},
  \bibinfo{journal}{Phys. Rev. B} \textbf{\bibinfo{volume}{67}},
  \bibinfo{pages}{155327} (\bibinfo{year}{2003}).

\bibitem[{\citenamefont{Bratkovsky and Levanyuk}(2006)}]{Bratkovsky-06.1}
\bibinfo{author}{\bibfnamefont{A.~M.} \bibnamefont{Bratkovsky}}
  \bibnamefont{and} \bibinfo{author}{\bibfnamefont{A.~P.}
  \bibnamefont{Levanyuk}}, \bibinfo{journal}{Integr. Ferroelectr.}
  \textbf{\bibinfo{volume}{84}}, \bibinfo{pages}{3} (\bibinfo{year}{2006});
\bibinfo{author}{\bibfnamefont{ibid}}, 
  \bibinfo{journal}{Appl. Phys. Lett.}
  \textbf{\bibinfo{volume}{89}}, \bibinfo{pages}{253108} (\bibinfo{year}{2006}).

\bibitem[{\citenamefont{Meyer and Vanderbilt}(2002)}]{Meyer-02}
\bibinfo{author}{\bibfnamefont{B.}~\bibnamefont{Meyer}} \bibnamefont{and}
  \bibinfo{author}{\bibfnamefont{D.}~\bibnamefont{Vanderbilt}},
  \bibinfo{journal}{Phys. Rev. B} \textbf{\bibinfo{volume}{65}},
  \bibinfo{pages}{104111} (\bibinfo{year}{2002}).

\bibitem[{\citenamefont{Prosandeev and Bellaiche}(2007)}]{Prosandeev-07}
\bibinfo{author}{\bibfnamefont{S.}~\bibnamefont{Prosandeev}} \bibnamefont{and}
  \bibinfo{author}{\bibfnamefont{L.}~\bibnamefont{Bellaiche}},
  \bibinfo{journal}{Phys. Rev. B} \textbf{\bibinfo{volume}{75}},
  \bibinfo{pages}{172109} (\bibinfo{year}{2007}).


\bibitem[{\citenamefont{Landau and Lifshitz}(1935)}]{Landau-35}
\bibinfo{author}{\bibfnamefont{L.}~\bibnamefont{Landau}} \bibnamefont{and}
  \bibinfo{author}{\bibfnamefont{E.}~\bibnamefont{Lifshitz}},
  \bibinfo{journal}{Phys. Z. Sowjetunion} \textbf{\bibinfo{volume}{8}},
  \bibinfo{pages}{153} (\bibinfo{year}{1935}).

\bibitem[{\citenamefont{Kittel}(1946)}]{Kittel-46}
\bibinfo{author}{\bibfnamefont{C.}~\bibnamefont{Kittel}},
  \bibinfo{journal}{Phys. Rev.} \textbf{\bibinfo{volume}{70}},
  \bibinfo{pages}{965} (\bibinfo{year}{1946}).

\bibitem[{\citenamefont{Lai et~al.}(2006)\citenamefont{Lai, Ponomareva, Naumov,
  Kornev, Fu, Bellaiche, and Salamo}}]{Lai-06}
\bibinfo{author}{\bibfnamefont{B.-K.} \bibnamefont{Lai}}
  \bibnamefont{et~al.}, 
  \bibinfo{journal}{Phys. Rev. Lett.}
  \textbf{\bibinfo{volume}{96}}, \bibinfo{pages}{137602}
  (\bibinfo{year}{2006}).

\bibitem[{\citenamefont{Stengel and Spaldin}(2006)}]{Stengel-06}
\bibinfo{author}{\bibfnamefont{M.}~\bibnamefont{Stengel}} \bibnamefont{and}
  \bibinfo{author}{\bibfnamefont{N.~A.} \bibnamefont{Spaldin}},
  \bibinfo{journal}{Nature (London)} \textbf{\bibinfo{volume}{443}},
  \bibinfo{pages}{679} (\bibinfo{year}{2006}).

\bibitem[{\citenamefont{Lichtensteiger
  et~al.}(2005)\citenamefont{Lichtensteiger, Triscone, Junquera, and
  Ghosez}}]{Lichtensteiger-05}
\bibinfo{author}{\bibfnamefont{C.}~\bibnamefont{Lichtensteiger}},
  \bibinfo{author}{\bibfnamefont{J.-M.} \bibnamefont{Triscone}},
  \bibinfo{author}{\bibfnamefont{J.}~\bibnamefont{Junquera}}, \bibnamefont{and}
  \bibinfo{author}{\bibfnamefont{Ph.}~\bibnamefont{Ghosez}},
  \bibinfo{journal}{Phys. Rev. Lett.} \textbf{\bibinfo{volume}{94}},
  \bibinfo{pages}{047603} (\bibinfo{year}{2005}).

\bibitem[{\citenamefont{Lichtensteiger
  et~al.}(2007)\citenamefont{Lichtensteiger, Dawber, Stucki, Triscone, Hoffman,
  Yau, Ahn, Despont, and Aebi}}]{Lichtensteiger-07}
\bibinfo{author}{\bibfnamefont{C.}~\bibnamefont{Lichtensteiger}}
  \bibnamefont{et~al.}, 
  \bibinfo{journal}{Appl. Phys. Lett.} \textbf{\bibinfo{volume}{90}},
  \bibinfo{pages}{052907} (\bibinfo{year}{2007}).

\bibitem[{\citenamefont{Nagarajan et~al.}(2006)\citenamefont{Nagarajan,
  Junquera, He, Jia, Lee, Kim, Zhao, Ghosez, Rabe, Baik et~al.}}]{Nagarajan-06}
\bibinfo{author}{\bibfnamefont{V.}~\bibnamefont{Nagarajan}}
  \bibnamefont{et~al.}, 
  \bibinfo{journal}{J. Appl. Phys.} \textbf{\bibinfo{volume}{100}},
  \bibinfo{pages}{1} (\bibinfo{year}{2006}).

\bibitem[{\citenamefont{Neaton and Rabe}(2003)}]{Neaton-03}
\bibinfo{author}{\bibfnamefont{J.~B.} \bibnamefont{Neaton}} \bibnamefont{and}
  \bibinfo{author}{\bibfnamefont{K.~M.} \bibnamefont{Rabe}},
  \bibinfo{journal}{Appl. Phys. Lett.} \textbf{\bibinfo{volume}{82}},
  \bibinfo{pages}{1586} (\bibinfo{year}{2003}).

\bibitem[{\citenamefont{Tian et~al.}(2006)\citenamefont{Tian, Jiang, Pan,
  Haeni, Li, Chen, Schlom, Neaton, Rabe, and Jia}}]{Tian-06}
\bibinfo{author}{\bibfnamefont{W.}~\bibnamefont{Tian}}
  \bibnamefont{et~al.}, 
  \bibinfo{journal}{Appl. Phys. Lett.} \textbf{\bibinfo{volume}{89}},
  \bibinfo{pages}{092905} (\bibinfo{year}{2006}).

\bibitem[{\citenamefont{Stephenson and Elder}(2006)}]{Stephenson-06}
\bibinfo{author}{\bibfnamefont{G.~B.} \bibnamefont{Stephenson}}
  \bibnamefont{and} \bibinfo{author}{\bibfnamefont{K.~R.} \bibnamefont{Elder}},
  \bibinfo{journal}{J. Appl. Phys.} \textbf{\bibinfo{volume}{100}},
  \bibinfo{pages}{051601} (\bibinfo{year}{2006}).

\bibitem[{\citenamefont{Baldereschi et~al.}(1988)\citenamefont{Baldereschi,
  Baroni, and Resta}}]{Baldereschi-88}
\bibinfo{author}{\bibfnamefont{A.}~\bibnamefont{Baldereschi}},
  \bibinfo{author}{\bibfnamefont{S.}~\bibnamefont{Baroni}}, \bibnamefont{and}
  \bibinfo{author}{\bibfnamefont{R.}~\bibnamefont{Resta}},
  \bibinfo{journal}{Phys. Rev. Lett.} \textbf{\bibinfo{volume}{61}},
  \bibinfo{pages}{734} (\bibinfo{year}{1988}).

\bibitem[{\citenamefont{Junquera et~al.}(2007)\citenamefont{Junquera, Cohen,
  and Rabe}}]{Junquera-07}
\bibinfo{author}{\bibfnamefont{J.}~\bibnamefont{Junquera}},
  \bibinfo{author}{\bibfnamefont{M.~H.} \bibnamefont{Cohen}}, \bibnamefont{and}
  \bibinfo{author}{\bibfnamefont{K.~M.} \bibnamefont{Rabe}},
  \bibinfo{journal}{J. Phys.: Condens. Matter} \textbf{\bibinfo{volume}{19}},
  \bibinfo{pages}{213203} (\bibinfo{year}{2007}).

\bibitem[{\citenamefont{He and Vanderbilt}(2003)}]{He-03}
\bibinfo{author}{\bibfnamefont{L.}~\bibnamefont{He}} \bibnamefont{and}
  \bibinfo{author}{\bibfnamefont{D.}~\bibnamefont{Vanderbilt}},
  \bibinfo{journal}{Phys. Rev. B} \textbf{\bibinfo{volume}{68}},
  \bibinfo{pages}{134103} (\bibinfo{year}{2003}).

\bibitem[{Sco()}]{Scott-porto}
\bibinfo{note}{J. F. Scott, ``Dynamics of ferroelectric nanostructures",
  XXXI International symposium on dynamical properties of solids, September
  2007, Porto, Portugal}.

\end{thebibliography}
\end{document}